\documentclass[prl,amssymb,amsmath,amsfonts,superscriptaddress,twocolumn,showpacs,twocolumn]{revtex4}
\usepackage[normalem]{ulem}
\usepackage{graphicx} 
\usepackage[usenames]{color}
\usepackage{amsmath,amssymb}
\usepackage{wasysym}
\usepackage{gensymb}
\usepackage{bbold}
\usepackage{natbib}

\usepackage[usenames,dvipsnames]{xcolor}
\usepackage{lipsum}
\usepackage{color}

\begin{document} 
\title{Emergent hyperuniformity  in periodically-driven emulsions}

\author{Joost H. Weijs}
\affiliation{Laboratoire de Physique de l'\'Ecole Normale Sup\'erieure de Lyon, Universit\'e de Lyon, 46, all\'ee d'Italie, 69007 Lyon, France}
\author{Rapha\"el Jeanneret}
\affiliation{Department of Physics, University of Warwick, Coventry, CV4 7AL, United Kingdom}

\author{R\'emi Dreyfus}
\affiliation{Complex Assemblies of Soft Matter, CNRS-Solvay-UPenn UMI 3254, Bristol, Pennsylvania 19007-3624, USA}

\author{Denis Bartolo}
\affiliation{Laboratoire de Physique de l'\'Ecole Normale Sup\'erieure de Lyon, Universit\'e de Lyon, 46, all\'ee d'Italie, 69007 Lyon, France} 
\date{\today} 

\begin{abstract}
We report the emergence of large-scale hyperuniformity  in microfluidic emulsions. Upon periodic driving  confined emulsions undergo a first-order transition from a reversible to an irreversible dynamics. We evidence that this dynamical transition is accompanied by structural changes at all scales yielding macroscopic yet finite hyperuniform structures. Numerical simulations are performed to single out the very ingredients responsible for the suppression of density fluctuations. We  show that as opposed to equilibrium systems the long-range nature of the hydrodynamic interactions are not required for the formation of hyperuniform patterns, thereby suggesting a robust relation between reversibility and hyperuniformity which should hold in a broad class of periodically driven materials.

\end{abstract} 
\maketitle

What is the most effective way to homogeneously fill space with an ensemble of particles? At thermal equilibrium, an obvious  effective strategy would be to endow the particles with interactions promoting the formation of a crystal. In a $d$-dimensional system, thermal fluctuations  would spontaneously organize the particles  into an ordered state  where the fluctuations  of the number of particles enclosed in a box of size $\ell$ would scale as $\Delta N^2_\ell\sim\ell^{d-1}$:  crystals are hyperuniform~\cite{TorquatoPRE03}. At large scales, they are much more homogeneous than  a random set of points with  number fluctuations of the order of the box volume $\ell^{d}$. However crystals are not the only patterns being hyperuniform~\cite{TorquatoPRE03,ZacharyJStatMech2009}. Over the last decade much attention has been devoted to disordered structures displaying miniature density fluctuations. As it turns out such hyperuniform patterns have been shown to display outstanding optical properties such as complete photonics band gaps~\cite{FlorescuPNAS2009,ChaikinOpExpr2013,Man2013}. Until very recently the only two controlled strategies to engineer hyperuniform materials were based on numerical optimization techniques~\cite{Batten2008,FlorescuPNAS2009,Man2013}, or  the jamming of athermal hard spheres~\cite{Donev2005,Zachary2011,Berthier2011,Weeks2010,Dreyfus}.  In 2015, two sets of numerical simulations have demonstrated that hyperuniformity emerges when ensembles of particles driven out of equilibrium approach  a critical absorbing phase transition~\cite{LevinePRL15,Berthier}. However, these numerical models based on  elegant toy models lack a truly analogous physical system, in which hyperuniformity emerges from genuine physical interactions.

In this letter we demonstrate experimentally that, when periodically driven, microfluidic emulsions self-organize into macroscopic (yet finite) hyperuniform structures. {\color{black} We evidence that maximal hyperuniformity is reached at the onset of reversibility of the droplet dynamics, even though the emulsion does not display any critical feature and does not reach a genuine absorbing state either.
We  also identify the minimal ingredients required to produce hyperuniform emulsions by means of numerical simulations. Surprisingly, unlike in equilibrium systems, we show that  long-range  interactions are not necessary to yield extended hyperuniform structures.} We therefore conjecture that hyperuniformity is intimately related to reversibility in periodically driven systems, and should therefore be achieved in a broad class of hard and soft condensed matter materials.

\begin{figure}
\begin{center}
  \includegraphics{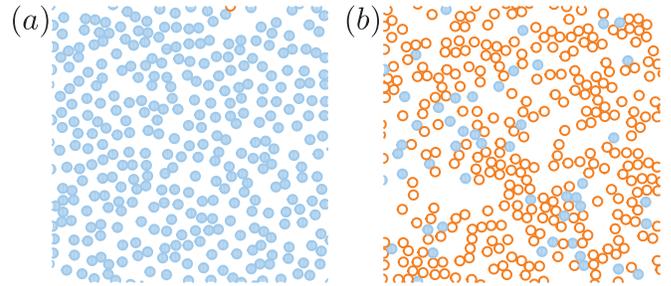}
  \caption{\label{fig:intro} Close-up of the emulsion at the beginning of the 500th cycle. The active droplets are shown as open orange  circles, the passive droplets as filled blue circles.(a) For a driving amplitude  $\Delta/\Delta^\star=0.65$, the dynamics is  reversible. (b) For a driving amplitude  $\Delta/\Delta^\star=1.54$, the dynamics is  not reversible anymore. Note also the markedly different structure in the two cases: for low amplitudes the structure is more homogeneous.}
%
%
\end{center}
\end{figure}
\begin{figure*}
\begin{center}
  \includegraphics{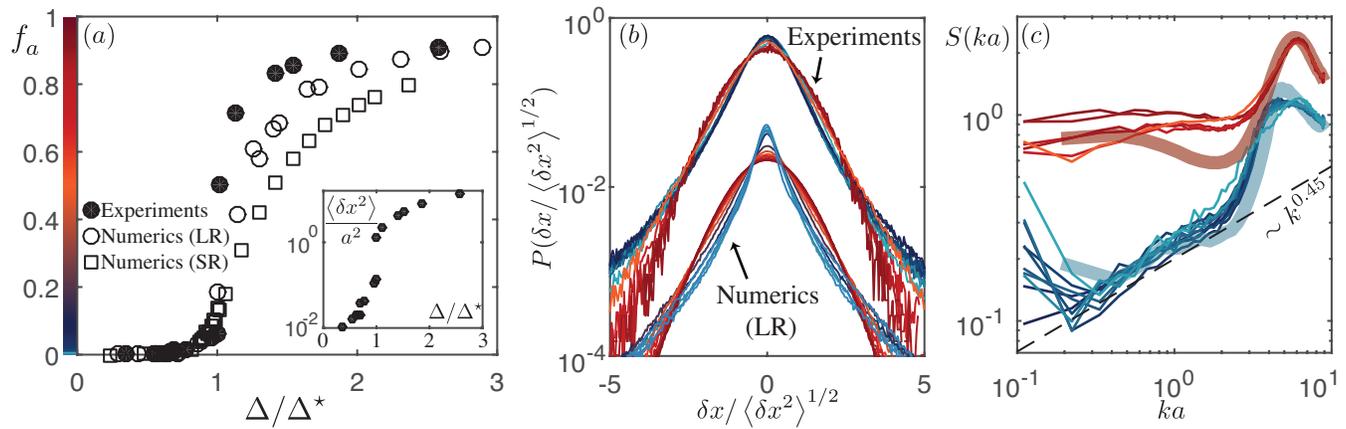}
  \caption{\label{fig:dynstructchanges}(a) Fraction of active particles $f_a$ in the steady state at various oscillation amplitudes. At $\Delta=\Delta^\star$ reversibility abruptly breaks down. Inset: Variations of the mean-squared strobed displacement in the flow direction $\langle\delta x^2\rangle$ plotted as a function of $\Delta$. (b) Centered and normalized probability density distribution of the strobed  displacements of the droplets, $\delta x$, in the flow direction for the experiments (top curves) and numerics (bottom curves). The numerical curves have been shifted for sake of clarity. The color indicates the mean fraction of active particles. See colorbar in (a).   (c) Structure factor $S(k)$ at various amplitudes (Experiments: thin lines, simulations: Thick lines, Same color code as in (b)). For sake of clarity the curves corresponding to $\Delta<\Delta^\star$ have been shifted down by a constant value. The dashed line is a guide and corresponds to $ k^{0.45}$.}
\end{center}
\end{figure*}
The experimental setup is the one used in~\cite{JeanneretNatCom2014} consisting of flowing a monodisperse emulsion (area fraction: 0.36) in a microfluidic channel (0.5\,cm$\times$5\,cm$\times 27\pm 0.1\,\mu$m). The droplets have a diameter $a=25.5\pm 0.5\,\mu$m comparable to the height of the channel and therefore undergo two-dimensional motion.
The inlet of the channel is connected to a syringe pump that drives the suspension sinusoidally.  The emulsion is prepared in a reproducible initial state reached after a sequence of 10 high-amplitude oscillations. Then, a sequence of $10^3$ cycles at the desired amplitude is applied to ensure that the measurements are performed in a statistically stationary state. 
The mean displacement of the droplets occurs along the main flow direction and is sinusoidal, and its amplitude $\Delta$ is the sole control parameter of the experiments. 
Data is collected by tracking the instantaneous position of $\sim 3\times10^3$ droplets remaining in the field of view throughout the entire flow cycle. 
Two snapshots of the emulsion  are shown in Fig.~\ref{fig:intro} and correspond to $\Delta=18.3a$ and $\Delta=43.3a$ respectively. 

Following~\cite{CorteNature2008} the macroscopic reversibility of the system is measured by determining the fraction of active particles $f_a$, which is the fraction of droplets that behave irreversibly. A droplet is here defined to be {\em active}  if it does not return at the end of a cycle within the spatial extent of the Vorono\"i cell it occupied at the start of the cycle. {\color{black}In these experiments $Re\ll 1$,  the fluid flows are therefore  reversible in time~\cite{JeanneretNatCom2014}. However, above a  driving amplitude $\Delta^*/a=28.1\pm0.3$ the droplet dynamics abruptly becomes irreversible. A macroscopic fraction of the droplets remains endlessly active upon periodic driving as illustrated in Fig.~\ 1, and quantified in Fig.~\ref{fig:dynstructchanges}a where $f_a$ is plotted as a function of driving amplitude. }
{\color{black}As the order parameter $f_a$ is based on a metric-free criteria, it is both  affected by the changes in the dynamics, and in the structure of the emulsion. 
We now disentangle and elucidate these two concomitant collective phenomena. }
Let us begin with the dynamical arrest of the strobed dynamics. The reversible states where droplets  retrace their steps back to their initial Vorono\"i cell, $\Delta<\Delta^\star$,  do not correspond to interaction-free conformations: the passive droplets continuously interact with their neighbors via hydrodynamic interactions in the course of the cycles. They 
also a priori experience a number of weak but irreversible 
perturbations such as short-range potential interactions, or minute shape deformations which {\color{black}cannot be  experimentally measured}. As a result, even in the reversible regime, they return only {\em on average} to their initial position after a cycle. The transition to an irreversible regime where all the particles are active is associated to a discontinuous amplification of the mean-square displacement of the strobed dynamics, thereby causing the escape of the droplets from their initial Vorono\"i cell, see Figure~\ref{fig:dynstructchanges}a inset.  We characterize the fluctuations of the strobed dynamics by the distribution of the  displacements at the end of each cycle in Figure~\ref{fig:dynstructchanges}b. At the onset of irreversibility this distribution is not merely widened, the statistics of the droplet displacements becomes qualitatively different at $\Delta^\star$ thereby confirming the discontinuous nature of the dynamical transition. Surprisingly, whereas the strobed-displacement statistics is Gaussian for high $\Delta$, it is exponential in the reversible regime, echoing the existence of rare large-amplitude jumps when the droplets exit their Vorono\"i cell. These intermittent displacements are akin to the cage jumps found in glass forming liquids~\cite{PRLberthierglass}. Importantly, these results imply that the
dynamical arrest of the strobed dynamics does not belong  to the absorbing-phase-transition scenario reported in~\cite{CorteNature2008,Sriram,LevinePRL15,Berthier}. 
\begin{figure*}
\begin{center}
  \includegraphics{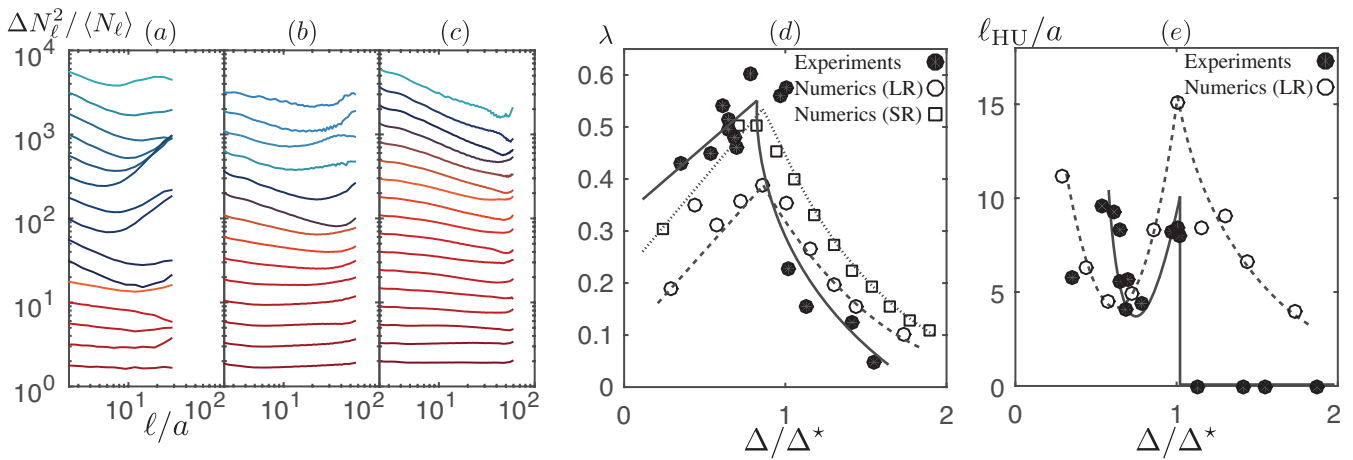}
  \caption{\label{fig:HU}(a), (b) and (c) Normalized density fluctuations for experiments (a), simulations with long-range interactions (b), and short-range interactions (c). At high amplitudes (red curves) the density fluctuations are normal, whereas in the reversible regime, at low amplitudes (blue curves), the density fluctuations are suppressed. (d) Fitted power-law exponents $\lambda$ from the curves plotted in (a), (b) and (c). The lines are guides to the eye. Density fluctuations are only suppressed for $\Delta\lesssim\Delta^\star$. (e) Extent of the hyperuniform regions $\ell_{HU}$. Filled symbols: experiments, open symbols: simulations.  The lines are guides to the eye. The system is only hyperuniform up to length-scales of $\sim10a$ (resp. $\sim15a$) in the experiments (resp. in the simulations).}
\end{center}
\end{figure*}

We now focus on the central results of this letter and demonstrate the emergence of hyperuniformity.
Having a quick look back at Fig.~\ref{fig:intro}, we see that the configuration in the reversible regime looks much more homogeneous than in the irreversible regime. In order to quantify this apparent structural change, 
we compute the structure factor $S(k)$ of the emulsion at various amplitudes,  Fig.~\ref{fig:dynstructchanges}c. The existence of a structural transition at the onset of reversibility is very noticeable from  the sudden change of $S(k)$ across all scales. The shift of of the high-$k$ peak of the structure factor corresponds to the change of the liquid-like structure reported in~\cite{JeanneretNatCom2014}. However the most striking feature at the transition occurs at low $k$. 
For high driving amplitudes (upper curves), $S(k)$  plateaus to a finite value as $k$ goes to 0, indicating the absence of long-range order. In contrast, at low amplitudes (lower curves) {\color{black}$S(k)$ decays algebraically as $k$ goes to 0, large-scale density fluctuations are suppressed~\cite{TorquatoPRE03}.} The emulsion self-organizes into a hyperuniform state.  

In order to quantify the degree of hyperuniformity and the extent of the hyperuniform regions, we go back to real space and directly measure the statistics of  the droplet number $N_\ell$ in a $\ell\times\ell$ box \cite{TorquatoPRE03,LevinePRL15}. Figure~\ref{fig:HU}a shows the variations of the variance $\Delta N^2_\ell\equiv \left<N_\ell^2\right>-\left<N_\ell\right>^2$ normalized by $\langle N_\ell\rangle$ as a function of $\ell$.
Any ensemble of particles with  no spatial correlation whatsoever has a variance that scales as $\Delta N^2_\ell\sim \left<N_\ell\right>$. 
Therefore, $\Delta N^2_\ell/\left<N_\ell\right>$ is a decreasing function of $\ell$ for hyperuniform systems. 
 In Figure~\ref{fig:HU}a we quantify the level of hyperuniformity  of the emulsions by normalizing the variance  $\Delta N^2_\ell$ by that of a random set of points of the same size and density. This procedure is used to minimize any statistical artifacts due to the finite sample size. For sake of clarity we have shifted the curves starting from small driving amplitudes at the top.
For large drivings $(\Delta\gg\Delta^\star)$, $\Delta N_\ell/\langle N_\ell\rangle$ does not show any significant variations. The density fluctuations are normal. Conversely,  as the dynamics becomes reversible  $(\Delta<\Delta^\star)$, the emulsion becomes locally hyperuniform. $\Delta N^2_\ell/N_\ell$ indeed first decays algebraically with $\ell$ up to a box size $\ell_{\rm HU}$ above which it increases. We expect the variations to then saturate for high values of $\ell$ that are not accessible with our experimental setup. This non-monotonic behavior is very similar to the one reported first in~\cite{LevinePRL15}, at the onset of an absorbing phase transition.

 More quantitatively,  the  system  homogeneity is quantified by the exponent $\lambda>0$ defined as: $\Delta N^2_\ell/\left<N_\ell\right>\propto \ell^{-\lambda}$. 
  For a random set of points $\lambda=0$, while $\lambda=1$ for perfect crystals. In Fig.~\ref{fig:HU}(b) the  exponent $\lambda$ fitted for small $\ell$ is shown as a function of the driving amplitude $\Delta$. Interestingly, 
the closer the system to the reversible transition, the more the density fluctuations are suppressed. At $\Delta=\Delta^\star$, we find that $\lambda \sim 0.5$ which is again  close to the one reported for systems close to an absorbing phase transition ($\lambda \sim 0.45$)  
 in~\cite{LevinePRL15} and in \cite{Berthier} at intermediate scales. 

The extent of the hyperuniform regions is characterized by measuring the length $\ell_{\rm HU}$ where $\Delta N^2_\ell/\left<N_\ell\right>$ deviates from a decreasing power law, Fig.~\ref{fig:HU}(c)~\cite{supmat}. This lengthscale undergoes non-monotonic variations with the driving amplitude.  When increasing $\Delta$, $\ell_{\rm HU}$ decays from ten to five  droplet diameters around $\Delta/\Delta^\star\sim0.7$. Then approaching the reversibility transition $\ell_{\rm HU}$ increases again to its maximal value ($\sim10a$) before dropping down to zero above $\Delta^\star$.  Two comments are in order. Firstly, we do not see any sign of a divergence of $\ell_{\rm HU}$ as $\Delta$ approaches $\Delta^\star$, which means that the emulsion never self-organizes into a fully hyperuniform state where the density fluctuations would be suppressed at the entire system scale. However, the typical extent of the hyperuniform regions are much larger than the typical distance  below which  the droplets display translational order. The pair correlation function of the emulsion decays exponentially over distances that are at most of the order of a couple of particle diameters~\cite{supmat}. Secondly, the rather complex variations of $\ell_{\rm HU}$ contrasts with that of all the other structural, and dynamical quantities which only display a significant change at the transition point  $\Delta^\star$. A potential explanation is that the intrinsic slowing down of the strobed dynamics below $\Delta^\star$ makes the hyperuniform self-organization  too slow to be experimentally achieved, although all the other (local) observables have reached a steady state.


What causes this emulsion to self-organize into hyperuniform large-scale structures? 
The droplets in the experimental system interact through various forces: hydrodynamic forces that are time reversible, but also short-range irreversible forces such  as depletion,  van der Waals, and electrostatic forces which are specific to the nature of the fluids and surfactants forming the emulsion.
To find out which of these ingredients are relevant to achieve hyperuniformity, we perform numerical simulations using a model containing only minimal hydrodynamic interactions and steric repulsion due to the finite size of the droplets.
The  flow induced by a moving droplet in a geometry as used in the experiment is described by a potential-flow dipole~\cite{BeatusPhysRep2012}. 
Each of the 896 droplets is advected by the local flow $\mathbf{u}(\mathbf{r})$ with the friction being modeled through a mobility coefficient $0<\mu<1$. 
The equation of motion for droplet $i$ is $\mathbf{\dot r_i}=\mu \mathbf{u}(\mathbf{r_i})$. 
Assuming pairwise additive interactions, the flow at the location of particle $i$ is $\mathbf{u}(\mathbf{r_i})$ is the sum of the contributions from the driving flow $\mathbf{u_0}$, and from the flow induced  $\mathbf{u_j}$ by all other particles $i\ne j$ in the system. 
The full equations of motion  are:
\begin{eqnarray}
\label{eq:eqmotion}
\mathbf{\dot r_i}=
\mu\left(\mathbf{u_0}(t)+ \sum_{j\ne i}{ \frac{2\mathbf{\hat{r}_{ij}}\mathbf{\hat{r}_{ij}}-\mathbb{1}}{2\pi |\mathbf{r_{ij}}|^2} \cdot \boldsymbol\sigma_j }\right)\;,
\end{eqnarray}
where $\boldsymbol\sigma_j$ is the dipole-vector associated with the particle $j$. The strength of the dipole is proportional to the  velocity of this particle relative to the ambient fluid. 
The boundary conditions in the flow-direction are periodic, whereas the flow is bounded by walls in the transverse direction. 
The infinite number of dipole-images that arise due to this are modeled analogous as in~\cite{SaintillanPRE2014} and is described in the supplementary materials~\cite{supmat}.

This minimal model  correctly accounts both for the dynamical and structural transitions. A reversible-to-irreversible transition occurs as the driving amplitude $\Delta$ is varied, Fig.~\ref{fig:dynstructchanges}a, open symbols.
The dipole strength which is our sole free parameter, is set to match the experimental value of $\Delta^\star$. Without any additional adjustment, we see that the same dynamical changes occur as in the experiments, Fig.~\ref{fig:dynstructchanges}b. The shape of the probability density abruptly changes from a Gaussian to an exponential across the reversibility transition as observed in the experiments.
Similarly, the structure above and below the transition are markedly different at all scales, Fig.~\ref{fig:dynstructchanges}c. Again the computed structure factor share the same salient features as the experimental one. 
Importantly, our simulations quantitatively captures the emergent hyperuniform structures in the reversible regime, Fig.~\ref{fig:HU}.
Both the exponent $\lambda$ and the extent $\ell_{HU}$ of  the hyperuniform regions are consistent with the experimental measurements, Fig.~\ref{fig:HU}b,c.
This agreement unambiguously  demonstrates (i) that hyperuniformity chiefly stems  from  the combination of reversible hydrodynamic interactions and short-range repulsion, and (ii)  that this phenomenology is robust to the very details of the interactions between the droplets and of their near-field  flows. 

In order to gain more physical insight, we now question the importance of the long-range ($\sim r^{-2}$ in 2D) nature of the interactions, which are very specific to hydrodynamics. 
Long-ranged interactions can yield hyperuniformity in systems at thermal equilibrium such as one-component plasmas~\cite{Lebovitz}. A natural question is therefore: does the emergence of hyperuniformity depend on the long-range nature of the particle-particle interactions in this non-equilibrium system as well?
To answer this question,  we perform the same simulations as above, but apply a very short-ranged exponential screening to the hydrodynamic interactions of the form $\exp(-|\mathbf{r_{ij}}|/a)$, keeping all the other parameters unchanged. 
Choosing a screening length of one particle diameter prevents the droplets from interacting over the observed hyperuniformity length scale while preserving the reversible nature of the microscopic dynamics. As is evident from Fig.~\ref{fig:dynstructchanges}a, the same reversible-to-irreversible transition, yet smoother, is observed
thereby further demonstrating the robustness of our main findings. Counterintuitively, the extent of the hyperuniform regions is clearly not reduced by screening, conversely it is extended up to the entire simulation window. As opposed to equilibrium systems, long-ranged interactions impair hyperuniform self-organization. This counterintuitive observation might be explained by a critical reversibility transition when the interactions are short-range, as suggested by the sharp increase of the structural relaxation time at $\Delta^\star$~\cite{supmat}. 
This picture is consistent with the results reported in~\cite{CorteNature2008} where hydrodynamic interactions are short-ranged and the transition critical.
However, a thorough finite-size scaling analysis would be required to unambiguously confirm that long-range interactions suppress criticality in our system.

{\color{black}
Together with that of~\cite{LevinePRL15,Berthier}, our experimental and numerical results strongly suggest that any ensemble of particles  at the onset of a reversible-to-irreversible transition  self-organizes into  hyperuniform  patterns.
A broad class of materials are therefore expected to self-assemble into hyperuniform structures upon periodic driving, from colloidal suspensions~\cite{PineNature2005,CorteNature2008,Nagel,Arratia}, to soft glasses~\cite{Ganapathy}, to shaken grains~\cite{Slotterback2012} to vortices in superconductors~\cite{Okuma2011}. }


\begin{acknowledgments}
We thank L. Berthier, P. Chaikin and S. Torquato  for insightful discussions. We acknowledge support support from Institut Universitaire de France (D. B.), and the NWO-Rubicon programme  financed by the Netherlands Organisation for Scientific Research (JHW).  
\end{acknowledgments}


\end{document}